\documentclass[conference]{IEEEtran}
\IEEEoverridecommandlockouts

\usepackage{cite}
\usepackage{amsmath,amssymb,amsfonts}
\usepackage{algorithmic}
\usepackage{graphicx,subfigure,multirow}
\usepackage{textcomp}
\usepackage{xcolor}
\usepackage{flushend}
\usepackage{fancyhdr}
\def\BibTeX{{\rm B\kern-.05em{\sc i\kern-.025em b}\kern-.08em
		T\kern-.1667em\lower.7ex\hbox{E}\kern-.125emX}}
	
\fancypagestyle{firstpage}{
	\setlength{\headheight}{3.5\normalbaselineskip}
	\setlength{\headsep}{1.5\normalbaselineskip}

	\fancyhead[C]{%
		\footnotesize%
		To be presented at the \textit{International Workshop on Multimedia Signal Processing} 2026, Istanbul, T\"urkiye, September 2026 \\
		\vspace{0.8\normalbaselineskip}%
		\scriptsize
		\copyright{} 2026 IEEE. Personal use of this material is permitted. Permission from IEEE must be obtained for all other uses, in any current or future media, including reprinting/republishing this material for advertising or promotional purposes, creating new collective works, for resale or redistribution to servers or lists, or reuse of any copyrighted component of this work in other works.%
	}
	\fancyfoot{}
}
	
\begin{document}
	
	\title{Bit Allocation Transfer for Perceptual Quality Enhancement of Traditional Video Codecs}
	
	\author{\IEEEauthorblockN{Runyu Yang}
		\IEEEauthorblockA{\textit{School of Engineering Science} \\
			\textit{Simon Fraser University}\\
			Burnaby, BC, Canada \\
			runyuy@sfu.ca}
		\and
		\IEEEauthorblockN{{Ivan V. Baji\'c} \thanks{This work was supported in part by the SFU-Huawei Joint Lab on Visual Computing.}}
		\IEEEauthorblockA{\textit{School of Engineering Science} \\
			\textit{Simon Fraser University}\\
			Burnaby, BC, Canada \\
			ibajic@sfu.ca}
	}
	
	\maketitle
	\thispagestyle{firstpage}
	
	\begin{abstract}
		
		Traditional block-based video codecs, such as H.264/AVC, H.265/HEVC and H.266/VVC, rely on hand-crafted Rate-Distortion Optimization (RDO) processes that primarily minimize Mean Squared Error (MSE), which correlates poorly with human perceptual quality. While neural video compression methods can easily optimize perceptually aligned metrics like MS-SSIM, their high computational complexity limits practical deployment. This paper proposes a novel bit allocation transfer framework that bridges these two paradigms to enhance the perceptual quality of conventional video codecs. Specifically, we train a quantization step generation model using a perceptual loss within a neural video compression framework (DCVC-FM). The model takes the original frame and a motion-compensated prediction as input and outputs a quantization step map. We then derive a block-wise bit ratio from this map and convert it into a Quantization Parameter (QP) map for a traditional video codec. Experimental results on the HEVC B$\sim$D dataset demonstrate that our method achieves 20.20\%, 8.25\%, and 8.37\% bitrate savings in terms of MS-SSIM compared with the standard reference software JM-19.0, HM-16.20, and VTM-23.0, respectively, with additional gains when utilizing predicted frames. Our approach effectively transfers the implicit perceptual importance learned by neural video compression models to guide block-level bit allocation in traditional video codecs without modifying their core decoding syntax.
		
	\end{abstract}
	
	
	
	\section{Introduction}
	
	Contemporary video compression standards, such as H.264/AVC, H.265/High Efficiency Video Coding (HEVC) \cite{sullivan2012overview}, and H.266/Versatile Video Coding (VVC) \cite{bross2021developments}, all adhere to a block-based hybrid coding framework. A central challenge in designing an encoder for any of these standards is achieving an optimal balance between compression efficiency (bitrate) and reconstruction quality. Formally, this is known as the Rate-Distortion Optimization (RDO) problem:
	\begin{equation}\label{equ:RDO}
		\min D \quad \text{subject to} \quad R \leq R_c,
	\end{equation}
	where \(D\) denotes the distortion (inversely related to quality), \(R\) is the actual bitrate consumed, and \(R_c\) represents the target bitrate constraint. Block-based coding inherently allows fine-grained RDO decisions at the block level. For each block, the encoder must make critical choices, including selecting the optimal prediction mode, determining the best partition size, and choosing an appropriate Quantization Parameter (QP) value.
	
	\begin{figure}
		\begin{minipage}[b]{1.0\linewidth}
			\centering
			\centerline{\includegraphics[width=\linewidth]{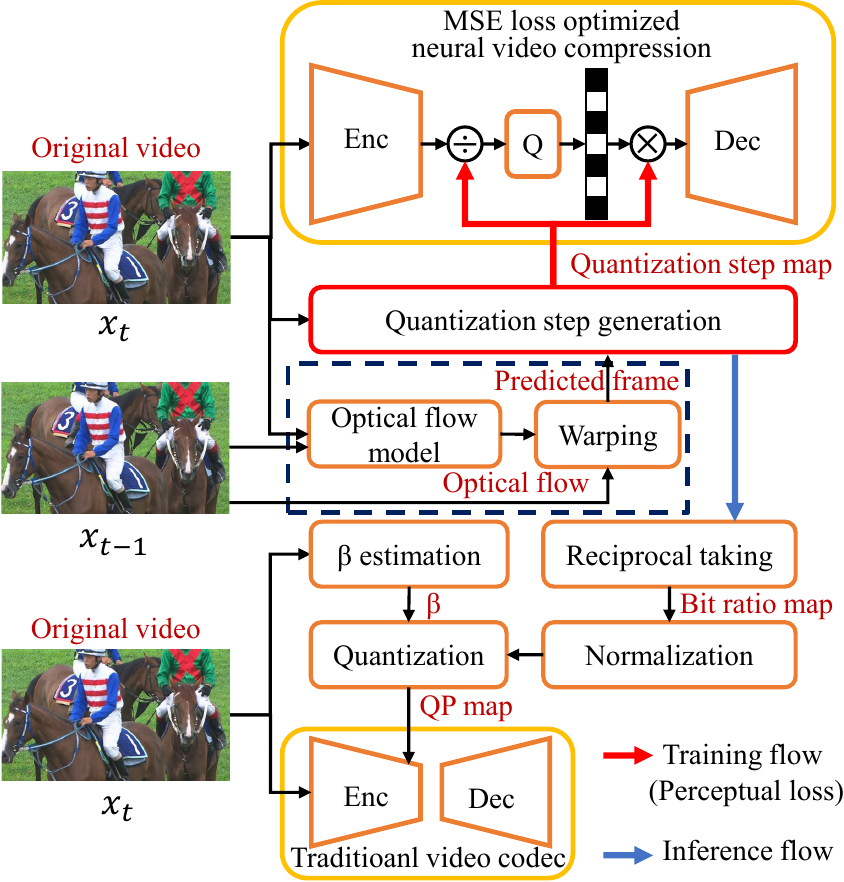}}
		\end{minipage}
		\caption{The proposed video coding framework with perceptual quality enhancement via transferring from neural video compression bit allocation. Note that the modules inside the dotted box are not used in I-frame coding.}
		\label{fig:overall}
	\end{figure}
	
	Crucially, the formulation and evaluation of distortion ($D$) play a decisive role in determining optimization performance. Mean Squared Error (MSE) and its logarithmic counterpart, Peak Signal-to-Noise Ratio (PSNR), have long been the dominant quality metrics because of their mathematical simplicity and analytical convenience. Nevertheless, extensive psychovisual studies have demonstrated that MSE/PSNR exhibits weak correlation with human perceptual judgments of visual quality \cite{girod1993s,wang2009mean}. To address this limitation, a wide range of perceptually-oriented quality metrics have been proposed to better capture characteristics of the human visual system. Representative examples include the Structural Similarity Index (SSIM) \cite{wang2004image}, the multi-scale variant MS-SSIM \cite{wang2003multiscale}, and learning-based perceptual metrics such as the Learned Perceptual Image Patch Similarity (LPIPS) \cite{zhang2018unreasonable}. Although these perceptual metrics consistently outperform MSE in terms of subjective quality correlation, they generally lack the additive decomposability and block-wise independence properties that make MSE particularly suitable for conventional RDO optimization. As a result, incorporating complex perceptual objectives such as MS-SSIM or LPIPS into block-level RDO optimization introduces substantial computational challenges and is often incompatible with the standard optimization framework, thereby limiting their practical deployment despite their superior perceptual fidelity.
	
	\begin{figure}
		\begin{minipage}[b]{1.0\linewidth}
			\centering
			\centerline{\includegraphics[width=\linewidth]{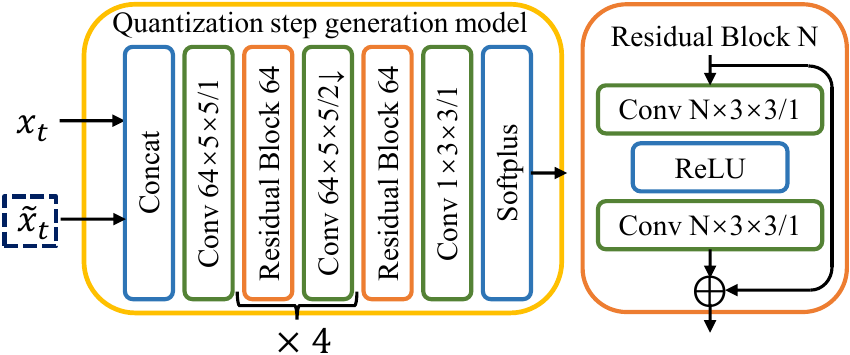}}
		\end{minipage}
		\caption{Left: Quantization step generation model; Right: Residual Block. The predicted frame $\tilde{x}_t$ inside the dotted box is not used in I-frame coding.}
		\label{fig:ada_model}
	\end{figure}
	
	Recently, neural video compression has emerged as a compelling alternative to traditional video coding frameworks~\cite{li2023neural, li2024neural, liao2025ehvc}. Unlike conventional codecs, this paradigm relies on deep neural networks to jointly model the entire encoder--decoder pipeline in an end-to-end manner. Consequently, its rate-distortion optimization (RDO) strategy differs fundamentally from that of classical approaches: optimization is performed globally during offline training over large-scale video datasets rather than locally at the block level during encoding. 
	
	One notable advantage of neural compression frameworks is their flexibility in distortion modeling. Any distortion metric that is differentiable with respect to the network parameters can be directly integrated into the gradient-based optimization process. This property enables perceptually-oriented objectives, such as MS-SSIM or LPIPS, to be incorporated far more naturally and efficiently than in conventional codecs. Furthermore, neural networks trained under different optimization criteria often produce substantially different spatial bit-allocation behaviors for the same input video. These learned allocation patterns implicitly encode the model's understanding of the relative perceptual significance of different spatial regions under the target quality metric.
	
	Previous works~\cite{yang2021knowledge,yang2024perceptual, yang2025bit} have demonstrated the implicit knowledge of perceptual block importance learned by learned image compression models can be extracted and transferred to guide RDO decisions within traditional block-based image codecs. This paper expands upon that foundational concept, presenting a bit allocation transfer framework for perceptual quality enhancement of traditional video codecs.
	
	The rest of this paper is organized as follows. Section \ref{sec:Method} describes the proposed perceptual quality enhancement framework. Section \ref{sec:Experiments} provides the detailed experimental results. Finally, Section \ref{sec:Conclusion} concludes this paper.
	
	\section{Proposed method}
	\label{sec:Method}
	
	The overall framework of our method is introduced in Section \ref{sec:Overall}, followed by detailed descriptions of various components in subsequent sections.
	
	\subsection{Overall framework}
	\label{sec:Overall}
	
	Fig.~\ref{fig:overall} presents the overall bit allocation transfer framework. Our goal is to generate a QP map for block-based video compression standards such as VVC in order to enhance perceptual quality. First, we train a quantization step generation model using a perceptual loss. The model takes as input the original frame together with the predicted frame, which is derived from warping the previous frame according to the estimated optical flow, and outputs a quantization step map. The reciprocal of this map serves as an estimate of the block-wise bit allocation ratio. After normalization for rate alignment, a bit ratio map is obtained. Subsequently, the target QP map is derived from the bit ratio map based on the R-$\lambda$ model \cite{li2014lambda}. Finally, the conventional video codec utilizes the generated QP map to perform adaptive bit allocation for perceptual quality enhancement.
	
	
	\subsection{Neural video compression}
	
	Our framework is model-agnostic and only requires the underlying neural compression architecture to support block-wise bit allocation. In this work, we adopt DCVC-FM for implementation \cite{li2024neural}. The model is optimized using the following joint rate-distortion objective:
	\begin{equation}\label{equ:loss function}
		L = \lambda D + R.
	\end{equation}
	Here, $\lambda$ balances the trade-off between bitrate and distortion, where $D$ denotes the distortion term measured by MSE or perceptual metrics, and $R$ represents the bitrate associated with the quantized latent representations and hyperpriors. The hyperpriors provide side information for entropy coding. The spatial resolution of the latent features is downsampled by a factor of 16 with respect to the original frame resolution. In addition, we directly reuse the optical flow model from DCVC-FM within our proposed framework.
	
	\subsection{Quantization step generation}
	
	The bitrate required for encoding quantized latent features is closely related to their magnitudes, where larger feature values generally correspond to higher bitrates, similar to the behavior of quantized coefficients in VVC. To achieve adaptive bit allocation, we introduce a quantization step generation model that outputs a quantization step map. This map inversely controls the magnitudes of the quantized latent features: values greater than one suppress the feature magnitudes and reduce bitrate consumption, whereas values smaller than one increase the feature magnitudes and increase the bitrate.
	
	As illustrated in Fig.~\ref{fig:ada_model}, the proposed model consists of stacked residual blocks and convolutional layers with stride 2. The output resolution matches that of the latent features. To guarantee positive outputs suitable for quantization step scaling, we adopt the softplus function as the final activation function, which is defined as
	\begin{equation}\label{equ:softplus}
		\text{softplus}(x) = \ln(1+e^x).
	\end{equation}
	
	During training, the parameters of the neural video compression model are kept fixed, while only the quantization step generation model is optimized. Two separate quantization step generation models are trained and applied to I-frames and P-frames, respectively.
	
	\begin{figure}
		\centering
		\subfigure[BasketballDrill]{
			\begin{minipage}[b]{0.48\linewidth}
				\includegraphics[width=1\linewidth]{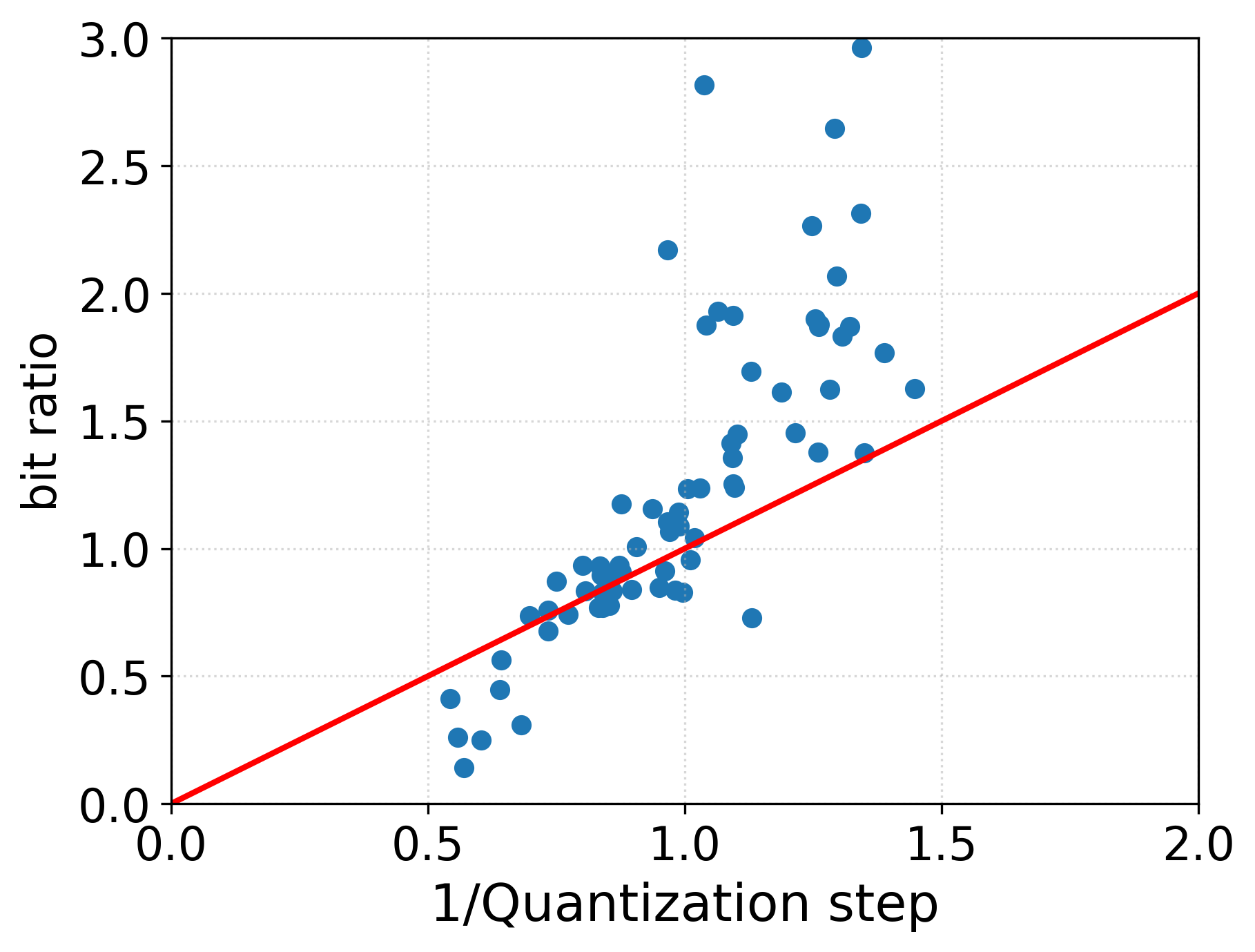}
		\end{minipage}}
		\subfigure[BQMall]{
			\begin{minipage}[b]{0.48\linewidth}
				\includegraphics[width=1\linewidth]{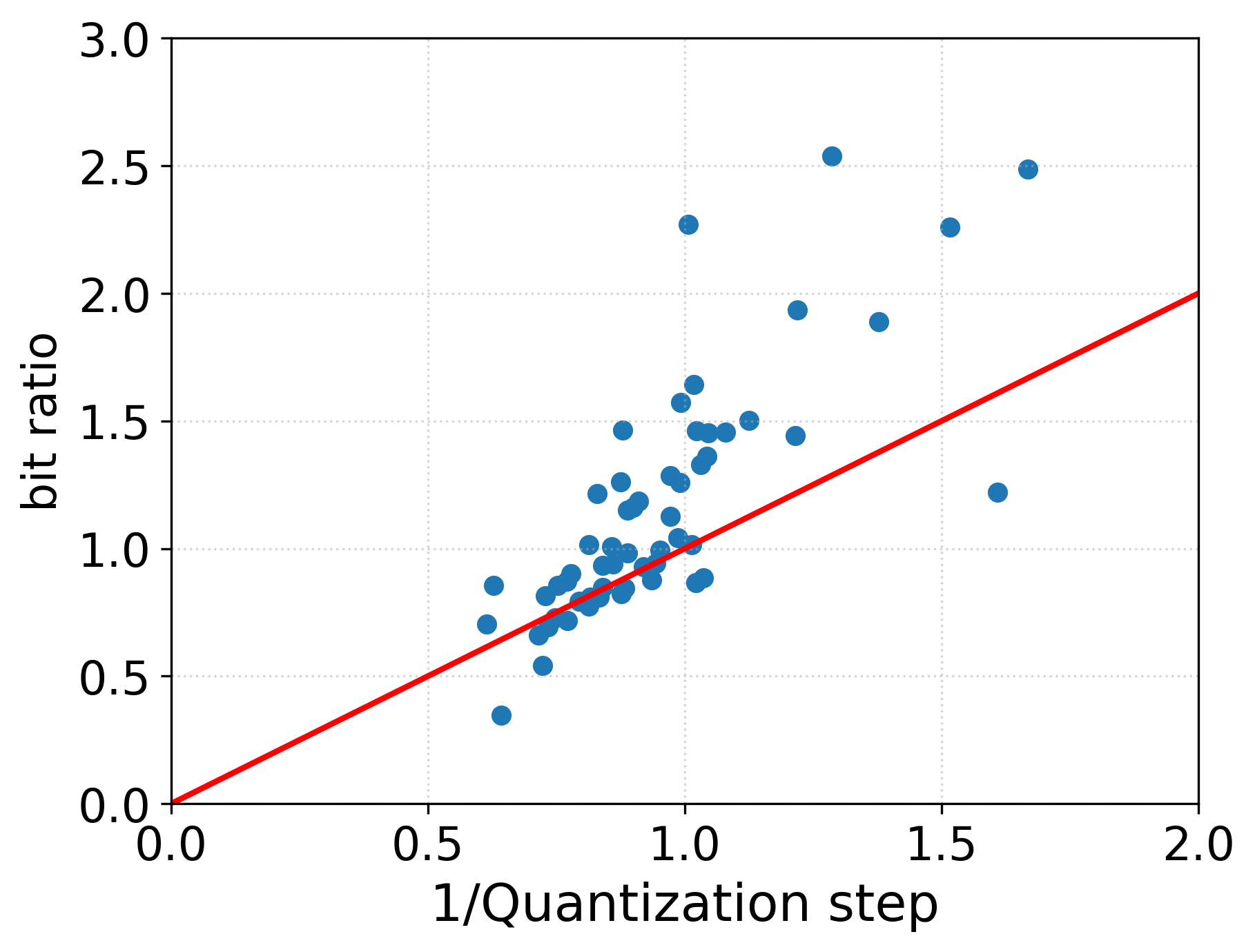}
		\end{minipage}}
		\caption{The relationship between bit ratio and the reciprocal of the quantization step on BasketballDrill and BQMall in Class C. The red line indicates a linear relationship with a slope of 1.}
		\label{fig:QS}
	\end{figure}
	
	\subsection{QP adaptation}
	
	To allocate bits adaptively for each block according to the quantization step map, the generated quantization steps must be converted into a corresponding QP map. We follow the strategy proposed in \cite{yang2024perceptual}, which relies on a block-wise bit allocation ratio. Empirically, we observe that the bit ratio is approximately proportional to the reciprocal of the quantization step as shown in Fig.~\ref{fig:QS} (granularity of 64$\times$64 block, take the average step of 4$\times$4). Accordingly, the bit ratio can be estimated as
	\begin{equation}\label{equ:ratio}
		r_k \approx \frac{1}{QS_k},
	\end{equation}
	where $k$ denotes the block index, $r_k$ represents the block-wise bit allocation ratio used for adaptive bit assignment, and $QS_k$ is the block-wise quantization step derived from averaging the quantization step values within the corresponding block region.
	
	To determine the bitrate $R$ from the Lagrange multiplier $\lambda$, we adopt the R-$\lambda$ model \cite{li2014lambda}, which is formulated as
	\begin{equation}\label{equ:R-lambda}
		\lambda = \alpha \cdot R^\beta,
	\end{equation}
	where $\alpha$ and $\beta$ are content-dependent hyperparameters. In the original formulation, both $\lambda$ and $R$ are defined at the frame level. Since our method performs QP adaptation on a block basis and the parameters $\alpha$ and $\beta$ vary across blocks, the model can be extended to
	\begin{equation}\label{equ:R-lambda2}
		\lambda = \alpha_k \cdot R_k^{\beta_k},
	\end{equation}
	where $k$ indicates the block index.
	
	To incorporate the target block-wise bit ratio $r_k$, the corresponding block-wise Lagrange multiplier $\lambda_k$ can be derived as
	\begin{align}\label{equ:R-lambda3}
		\lambda_k &= \alpha_k \cdot \left(r_k R_k\right)^{\beta_k}\\
		&= r_k^{\beta_k} \cdot \alpha_k \cdot R_k^{\beta_k}\\
		&= r_k^{\beta_k} \cdot \lambda.
	\end{align}
	
	The parameter $\beta_k$ is estimated using the neural network proposed in \cite{li2017convolutional}. Based on the relationship between QP and $\lambda$, the QP value for the $k$-th block can be expressed as
	\begin{align}\label{equ:QP2}
		\text{QP}_k &= \text{QP} + N\log_2 \left(\frac{\lambda_k}{\lambda}\right) \\
		&= \text{QP} + 3\log_2 \left(r_k^{\beta_k}\right)\\
		&\approx \text{QP} - 3\log_2 \left(QS_k^{\beta_k}\right),
	\end{align}
	where $N$ is set to 3 following the AVC, HEVC, and VVC configuration. The corresponding block-wise Lagrange multiplier $\lambda_k$ is then updated as
	\begin{equation}\label{equ:lambda}
		\lambda_k = \lambda \cdot 2^{(\text{QP}_k-\text{QP})/3}.
	\end{equation}
	
	\section{Experiments}
	\label{sec:Experiments}
	
	\subsection{Settings}
	
	The training strategy and network architecture for $\beta_k$ estimation follow the same settings as those in \cite{yang2024perceptual}. The training data for the quantization step generation model is derived from the Vimeo-90k dataset \cite{xue2019video}, which contains 90,000 randomly cropped patches with a resolution of 192$\times$192.
	
	We evaluate our method using JM-19.0, HM-16.20, and VTM-23.0, which are the reference software implementations for AVC, HEVC, and VVC, respectively. The QP adaptation block size is configured as 16$\times$16 for JM-19.0, and 64$\times$64 for both HM-16.20 and VTM-23.0. For comparison with PerceptQPA \cite{helmrich2018Improved} in VTM-23.0, we replace its original block-wise QP adaptation module while retaining its chroma-aware bit allocation mechanism. To alleviate potential blocking artifacts, the maximum QP offset is constrained to 5.
	
	We adopt $1-\text{MS-SSIM}$ as the perceptual loss function during training. The neural video compression model is initialized using pre-trained MSE-optimized weights, while the quantization step generation model is trained for five epochs using the Adam optimizer with an initial learning rate of $1\times10^{-4}$.
	
	The evaluation metrics include RGB PSNR and MS-SSIM. Experiments are conducted on the standard HEVC Class B$\sim$D test sequences under low-delay coding configurations \cite{bossen2013Common}.
	
	\subsection{Results}
	
	Tables~\ref{table:result1}--\ref{table:result3} present the BD-rate performance of the proposed method on AVC, HEVC, and VVC reference codecs, respectively. The results demonstrate that the proposed perceptual bit allocation strategy consistently improves perceptual quality in terms of MS-SSIM while maintaining competitive PSNR performance.
	
	For AVC in Table~\ref{table:result1}, the proposed method achieves an average MS-SSIM BD-rate reduction of 20.20\% with only a 7.67\% BD-rate increase measured in terms of PSNR relative to JM-19.0. Compared with the variant without the predicted frame, incorporating the predicted frame generally leads to further perceptual gains, improving the overall MS-SSIM BD-rate reduction from 18.36\% to 20.20\%. The improvement is particularly evident on sequences with strong temporal correlations, such as \textit{BQTerrace} and \textit{BQSquare}. These results indicate that the predicted frame provides useful temporal guidance for perceptual bit allocation. In addition, the proposed method achieves substantial perceptual improvements across all classes, demonstrating good robustness under different spatial resolutions and motion characteristics.
	
	Table~\ref{table:result2} reports the experimental results on HM-16.20. The proposed method achieves an overall MS-SSIM BD-rate reduction of 8.25\% while introducing only a modest PSNR BD-rate increase of 4.88\%. Compared with AVC, the gains on HEVC are relatively smaller because HEVC already includes more advanced coding tools and larger QP adaptation block size. Nevertheless, the proposed method still consistently improves perceptual quality across most sequences. Incorporating the predicted frame again provides additional benefits in several sequences. For example, the MS-SSIM BD-rate reduction on \textit{BQTerrace} improves from 16.97\% to 17.52\% after introducing temporal prediction information. These results confirm that the proposed quantization step generation model remains effective even in modern hybrid coding frameworks.
	
	Table~\ref{table:result3} compares the proposed method with PerceptQPA on VTM-23.0. Because perceptQPA does not support block-wise QP adaptation for 480p, we evaluate only on Class B and Class C. PerceptQPA achieves an overall MS-SSIM BD-rate reduction of 6.98\%, whereas the proposed method further improves the performance to 8.37\%. In contrast, the PSNR increase of the proposed method remains comparable to that of PerceptQPA. This demonstrates that the proposed framework achieves perceptual quality enhancement without causing excessive degradation in distortion-oriented quality metrics. Moreover, compared with the variant without the predicted frame, the full model consistently achieves slightly better MS-SSIM performance, validating the effectiveness of incorporating temporal prediction information into the quantization step generation process. Overall, the experimental results verify that the proposed method generalizes well across different coding standards and provides superior perceptual bit allocation performance compared with existing QP adaptation approaches.
	
\begin{figure*}
		\begin{minipage}[b]{1.0\linewidth}
			\centering
			\centerline{\includegraphics[width=\linewidth]{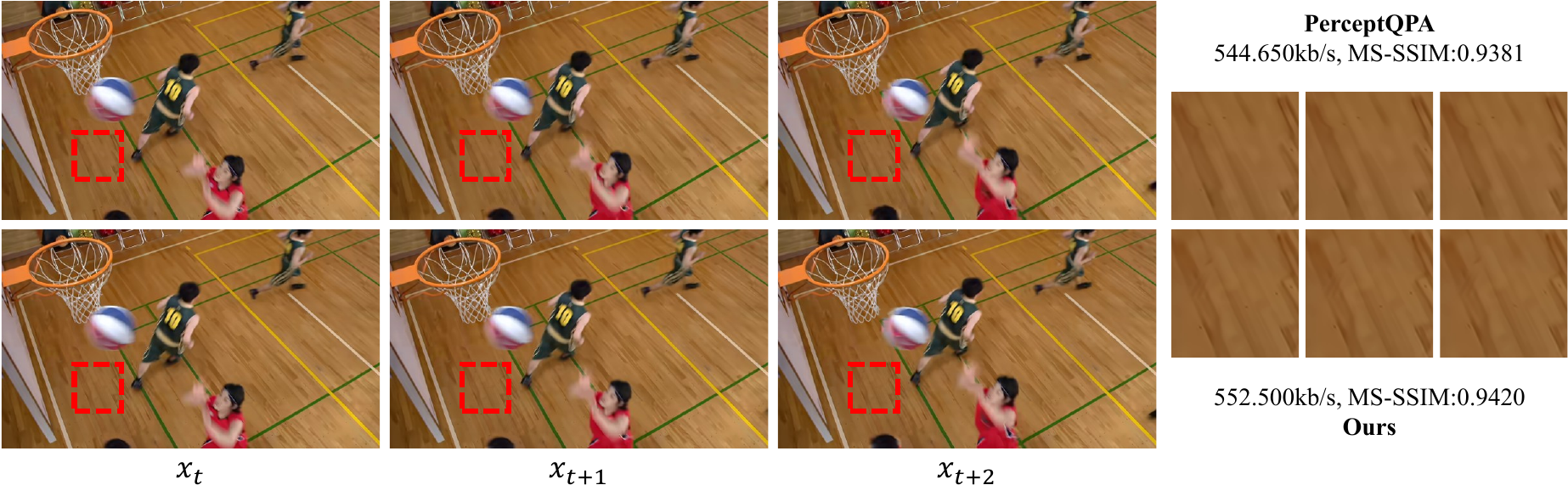}}
		\end{minipage}
		\caption{Visual results of reconstruction frames in our method and perceptQPA.}
		\label{fig:visual}
	\end{figure*}

	Visual comparisons between our method and PerceptQPA implemented in VTM-23.0 are shown in Fig.~\ref{fig:visual}. We can observe that our method decreases the QP of motionless and low-texture regions which can improve the perceptual quality without significant increase in bitrate.
	
	Regarding computational complexity, we measured the runtime of the proposed QP map generation on a desktop PC running Windows 11 (64-bit), equipped with AMD Ryzen 7 5800H CPU, 16 GB RAM, and an NVIDIA GeForce GTX 1080 Ti GPU, using PyTorch 2.0.0. For a single 1080p frame, the proposed method requires approximately 0.2 s on the GPU and 5.3 s on the CPU. This overhead represents only a small fraction of the overall VVC encoding time, which exceeds 1 min per frame at QP 37.
	
	\begin{table}
		\centering
		\caption{BD-rate results (\%) of our method and our method without the predicted frame, compared to JM-19.0}
		
		\begin{tabular}{cc|cc|cc}
			\hline
			\multirow{2}{*}{Class}&\multirow{2}{*}{Sequence}&\multicolumn{2}{c|}{Ours w/o pred. frame}&\multicolumn{2}{c}{Ours}\\ \cline{3-6}
			&& PSNR & MS-SSIM & PSNR & MS-SSIM\\
			\hline
			\multirow{3}{*}{B}&BasketballDrive &6.16 &$-$12.42 &6.17 &$-$14.17 \\
			&BQTerrace &11.70 &$-$36.86 &11.09 &$-$40.45 \\
			&Cactus&17.91 &$-$12.30 &22.34 &$-$13.66\\
			\hline
			\multirow{4}{*}{C}&BasketballDrill &4.94 &$-$18.99 &4.42 &$-$20.73 \\
			&BQMall &8.53 &$-$13.01 &9.36 &$-$14.24 \\
			&PartyScene &9.87 &$-$10.07 &10.79 &$-$9.95 \\
			&RaceHorses	&4.44 &$-$12.98 &5.14 &$-$13.78 \\
			\hline
			\multirow{4}{*}{D}&BasketballPass &$-$6.63 &$-$27.52 &$-$7.54 &$-$28.83\\
			&BlowingBubbles &5.32 &$-$13.63 &7.55 &$-$16.61 \\
			&BQSquare &4.30 &$-$31.73 &2.53 &$-$37.17 \\
			&RaceHorses	&7.68 &$-$10.33 &7.00 &$-$9.97 \\
			\hline
			\multicolumn{2}{c|}{Class B}&11.92 &$-$20.53 &13.20 &$-$22.76 \\
			\multicolumn{2}{c|}{Class C}&6.95 &$-$13.76 &7.43 &$-$14.68 \\
			\multicolumn{2}{c|}{Class D}&2.67 &$-$20.80 &2.39 &$-$23.15 \\
			\hline
			\multicolumn{2}{c|}{Overall}&7.18 &$-$18.36 &7.67 &$-$20.20 \\
			\hline
			
		\end{tabular}
		\label{table:result1}
	\end{table}
	
	\begin{table}
		\centering
		\caption{BD-rate results (\%) of our method and our method without the predicted frame, compared to HM-16.20}
		
		\begin{tabular}{cc|cc|cc}
			\hline
			\multirow{2}{*}{Class}&\multirow{2}{*}{Sequence}&\multicolumn{2}{c|}{Ours w/o pred. frame}&\multicolumn{2}{c}{Ours}\\ \cline{3-6}
			&& PSNR & MS-SSIM & PSNR & MS-SSIM\\
			\hline
			\multirow{3}{*}{B}&BasketballDrive &4.76 &$-$8.33 &6.83 &$-$6.92 \\
			&BQTerrace &9.51 &$-$18.67 &7.97 &$-$20.36 \\
			&Cactus&8.68 &$-$7.54 &10.14 &$-$8.37\\
			\hline
			\multirow{4}{*}{C}&BasketballDrill &8.67 &$-$8.74 &8.15 &$-$9.20 \\
			&BQMall &3.89 &$-$5.04 &4.03 &$-$5.17 \\
			&PartyScene &3.56 &$-$3.50 &3.82 &$-$3.59 \\
			&RaceHorses	&2.63 &$-$4.32 &1.88 &$-$4.17 \\
			\hline
			\multirow{4}{*}{D}&BasketballPass &$-$2.10 &$-$11.69 &$-$2.29 &$-$12.92\\
			&BlowingBubbles &2.14 &$-$0.90 &2.82 &$-$1.54 \\
			&BQSquare &4.25 &$-$9.37 &4.11 &$-$11.36 \\
			&RaceHorses	&4.42 &$-$3.51 &2.84 &$-$3.49 \\
			\hline
			\multicolumn{2}{c|}{Class B}&7.65 &$-$11.51 &8.31 &$-$11.88 \\
			\multicolumn{2}{c|}{Class C}&4.69 &$-$5.40 &4.47 &$-$5.53 \\
			\multicolumn{2}{c|}{Class D}&2.18 &$-$6.37 &1.87 &$-$7.33 \\
			\hline
			\multicolumn{2}{c|}{Overall}&4.84 &$-$7.76 &4.88 &$-$8.25 \\
			\hline
			
		\end{tabular}
		\label{table:result2}
	\end{table}

	\begin{table*}
		\centering
		\caption{BD-rate results (\%) of our method, our method without the predicted frame and PerceptQPA, compared to VTM-23.0}
		
		\begin{tabular}{cc|cc|cc|cc}
			\hline
			\multirow{2}{*}{Class}&\multirow{2}{*}{Sequence}&\multicolumn{2}{c|}{PerceptQPA}&\multicolumn{2}{c|}{Ours w/o pred. frame}&\multicolumn{2}{c}{Ours}\\ \cline{3-8}
			&& PSNR & MS-SSIM & PSNR & MS-SSIM& PSNR & MS-SSIM\\
			\hline
			\multirow{3}{*}{B}&BasketballDrive &1.96&$-$7.84&3.03 &$-$7.42 &4.57 &$-$7.13 \\
			&BQTerrace &8.28&$-$15.87&9.84 &$-$16.97 &9.60 &$-$17.52 \\
			&Cactus&7.87&$-$6.41&7.96 &$-$5.97 &8.58 &$-$6.92\\
			\hline
			\multirow{4}{*}{C}&BasketballDrill &10.46&$-$6.55&5.82 &$-$8.67 &5.54 &$-$8.99 \\
			&BQMall &4.49&$-$6.29&2.49 &$-$6.98 &2.47 &$-$7.03 \\
			&PartyScene &3.95&$-$2.45&2.58 &$-$4.84 &2.90 &$-$4.99 \\
			&RaceHorses	&6.36&$-$0.34&1.42 &$-$4.11 &0.53 &$-$3.88 \\
			\hline
			\multicolumn{2}{c|}{Class B}&6.04&$-$10.04&6.94 &$-$10.12 &7.58 &$-$10.52 \\
			\multicolumn{2}{c|}{Class C}&6.32&$-$3.91&3.08 &$-$6.15 &2.86 &$-$6.22 \\
			\hline
			\multicolumn{2}{c|}{Overall}&6.18&$-$6.98&5.01 &$-$8.14 &5.22 &$-$8.37 \\
			\hline
			
		\end{tabular}
		\label{table:result3}
	\end{table*}
	

	\section{Conclusion}
	\label{sec:Conclusion}
	This paper presents a novel bit allocation transfer framework that bridges neural video compression and traditional video codecs to enhance perceptual quality. By extracting the implicit knowledge of perceptual importance learned by neural compression models—specifically, the block-wise bit allocation patterns optimized for perceptually aligned metrics like MS-SSIM—we generate quantization step maps that guide adaptive QP adaptation within conventional codecs such as H.266/VVC. Experimental results on the HEVC B$\sim$D dataset demonstrate that our approach achieves 20.20\%, 8.25\%, and 8.37\% bitrate savings in terms of MS-SSIM compared with the reference implementations JM-19.0, HM-16.20 and VTM-23.0, respectively, with additional gains when utilizing predicted frames. These findings confirm that perceptual optimization strategies from neural video compression can be effectively transferred to improve the rate-distortion performance of traditional video codecs. Future work will explore extending this framework to task-oriented optimization objectives.

	\bibliographystyle{IEEEtran}
	\bibliography{refs}

\end{document}